# Optical detection of bond-dependent and frustrated spin in the two-dimensional cobalt-based honeycomb antiferromagnet $Cu_3Co_2SbO_6$

Baekjune Kang,[1*] Uksam Choi,[1*] Taek Sun Jung,[2*] Seunghyeon Noh,[3*] Gye-Hyeon Kim,[1] UiHyeon Seo,[1] Miju Park,[1] Jin-Hyun Choi,[1] Minjae Kim,[4] GwangCheol Ji,[4] Sehwan Song,[4] Hyesung Jo,[5] Seokjo Hong,[5] Nguyen Xuan Duong,[6] Tae Heon Kim,[6] Yongsoo Yang,[5] Sungkyun Park,[4] Jong Mok Ok,[4] Jung-Woo Yoo,[3§] Jae Hoon Kim,[2‡] and Changhee Sohn[1†]

[1]Department of Physics, Ulsan National Institute of Science and Technology, Ulsan, 44919, Republic of Korea
[2]Department of Physics, Yonsei University, Seoul, 03722, Republic of Korea
[3]Department of Materials Science and Engineering, Ulsan National Institute of Science and Technology, Ulsan, 44919, Republic of Korea
[4]Department of Physics, Pusan National University, Pusan, 46241, Republic of Korea
[5]Department of Physics, Korea Advanced Institute of Science and Technology, Daejeon, 34141, Republic of Korea
[6]Department of Physics, University of Ulsan, Ulsan, 44610, Republic of Korea

**Abstract**

Two-dimensional honeycomb antiferromagnet becomes an important class of materials as it can provide a route to Kitaev quantum spin liquid, characterized by massive quantum entanglement and fractional excitations. The signatures of its proximity to Kitaev quantum spin liquid in the honeycomb antiferromagnet includes anisotropic bond-dependent magnetic responses and persistent fluctuation by frustration in paramagnetic regime. Here, we propose $Cu_3Co_2SbO_6$ heterostructures as an intriguing honeycomb antiferromagnet for quantum spin liquid, wherein bond-dependent and frustrated spins interact with optical excitons. This system exhibits antiferromagnetism at 16 K with different spin-flip magnetic fields between a bond-parallel and bond-perpendicular directions, aligning more closely with the generalized Heisenberg-Kitaev than the XXZ model. Optical spectroscopy reveals a strong excitonic transition coupled to the antiferromagnetism, enabling optical detection of its spin states. Particularly, such spin-exciton coupling presents anisotropic responses between bond-parallel and bond-perpendicular magnetic field as well as a finite spin-spin correlation function around 40 K, higher than twice its Néel temperature. The characteristic temperature that remains barely changed even under strong magnetic fields highlights the robustness of the spin-fluctuation region. Our results demonstrate $Cu_3Co_2SbO_6$ as a unique candidate for the quantum spin liquid phase, where the spin Hamiltonian and quasiparticle excitations can be probed and potentially controlled by light.

**Main**

While at least 16,300 materials have been discovered/proposed as magnetic,[1] fewer than 10 materials have been suggested as potential Kitaev quantum spin liquid (QSL) candidates.[2,3] Kitaev QSL represents a rare, exactly solvable ground state of the Kitaev two-dimensional honeycomb model,[4] retaining massive quantum entanglement without long-range ordering even at zero kelvin. Their excitations are described by fractionalized Majorana fermions and non-Abelian anyons, providing a promising avenue for achieving fault-tolerant quantum computation through the non-Abelian braiding process.[5] However, despite their intriguing properties, the realization of Kitaev QSL remains extremely challenging. This exotic model can be implemented in edge-shared octahedral systems with strong spin-orbit coupling.[6] Therefore, 4$d$/5$d$ transition metal systems, such as $\alpha$-RuCl$_3$[7-10] and Na$_2$IrO$_3$[11-13], have been proposed as possible hosts for Kitaev QSL. Despite the emergence of long-range antiferromagnetic ordering at low temperature ($T$), these candidate materials still exhibit signatures indicative of proximity to Kitaev QSL, including half-integer thermall Hall,[9] broad continuum in inelastic neturon[8] and Raman scattering[14], anisotropic bond-dependent responses[15] and high-$T$ spin fluctuations[16].

Recently, high-spin 3$d^7$ cobalt-based honeycomb materials have been suggested as potential Kitaev QSL candidates.[17-21] This proposal is based on the ground state characterized by $J_{eff}$ = 1/2, which comprises the total angular momentum $L$ = 1 and total spin momentum $S$ = 3/2 of high-spin 3$d^7$ electronic configuration despite its relatively smaller spin-orbit coupling compared to 4$d$/5$d$ transition metal oxides. Furthermore, the localized nature of Co 3$d$ orbitals might suppress undesired next-nearest neighbor and direct exchange interactions. Various experimental and theoretical studies have provided mixed conclusions on the feasibility of Kitaev QSL in these 3$d^7$ cobalt-based honeycomb systems. For instance, the antiferromagnetic 1/3 order, the $\Gamma$-$M$ characteristics of the magnetic ordering wave vectors, and giant in-plane magnetic anisotropy of Na$_3$Co$_2$SbO$_6$ are well described as bond-dependent Kitaev Hamiltonian with a positive off-diagonal term $\Gamma_1$.[15,22,23] Using time-domain terahertz spectroscopy on BaCo$_2$(AsO$_4$)$_2$, a broad magnetic continuum, a signature of fractionalization, has been observed.[24] Recent theoretical and experimental investigations, however, suggest that cobalt-cobalt direct hopping is too substantial to ignore, particularly in BaCo$_2$(AsO$_4$)$_2$,[25] leading to an isotropic XXZ model, not Kitaev Hamiltonian.[15] Therefore, it remains elusive whether a cobalt-based honeycomb antiferromagnet is proximate to Kitaev QSL or not.

Here, we propose $Cu_3Co_2SbO_6$ heterostructures as an intriguing QSL candidate wherein there's an interplay between bond-dependent/frustrated spins and optical excitons. $Cu_3Co_2SbO_6$ has an alternatively stacked structure of a $Cu^+$ layer and a $(Co_{2/3}Sb_{1/3}O_2)^-$ layer with honeycomb edge-sharing $CoO_6$ octahedra (Fig. 1a and Fig. S1). Although single-phase $Cu_3Co_2SbO_6$ does not exist in nature,[26] we achieved its synthesis through heterostructure epitaxy on both ZnO and $MgAl_2O_4$ substrates. The magnetic susceptibility $\chi(T)$ suggests an antiferromagnetic ground state below 16 K. A pronounced in-plane magnetic anisotropy between bond-parallel and bond-perpendicular directions was observed, particularly in the spin-flip magnetic field, indicating that the spin Hamiltonian of $Cu_3Co_2SbO_6$ is more consistent with the anisotropic generalized Heisenberg-Kitaev Hamiltonian rather than the isotropic XXZ model.[15] A distinctive characteristic of $Cu_3Co_2SbO_6$ compared to other Kitaev systems is the formation of excitons between $Cu^+$ and $(Co_{2/3}Sb_{1/3}O_2)^-$ layers and their strong interaction with antiferromagnetic ordering (Fig. 1a). Through optical spectroscopy, we identified a strong Cu $3d \rightarrow$ Co $3d$ transition-type exciton near 4 eV, exhibiting peculiar spin-exciton coupling. We observed clear anomaly not only at Néel temperature $T_N \sim$ 16 K, but also an additional anomaly at $T_H \sim$ 40 K in the raw ellipsometry parameter, intensity, scattering rate, and peak position of the excitonic peak. The occurrence of $T_H$, which is indicative of a finite spin-spin correlation function far above the $T_N$, implies strong frustrated exchange interactions, a key ingredient of QSL. The $T_H$ showed little magnetic field ($H$) dependence, underscoring the robustness of the spin fluctuation region. Based on $H$-dependent $\chi$ and optical spectroscopy, we constructed $T$-$H$ phase diagram (Fig. 1b), which highlights the presence of unconventional spin fluctuation regime between antiferromagnetic and conventional paramagnetic phases.

**Synthesis of single-crystalline, epitaxial $Cu_3Co_2SbO_6$ thin film**

We successfully synthesized single-phase $Cu_3Co_2SbO_6$, which does not naturally exist, via epitaxy on ZnO and $MgAl_2O_4$ substrates. During the synthesis of bulk samples, substantial magnetic impurities $(Co,Sb)_3O_4$ were inevitably formed due to the similarity between ordered stacking temperature (1250 °C) and decomposition temperature (1260 °C). This has hindered understanding the intrinsic properties for this Kitaev QSL candidate.[26,27] However, in thin film geometry, the epitaxial relationship with the substrate and its energetic synthesis mechanism can result in different consequences from bulk synthesis.[28] Single phase, epitaxial $Cu_3Co_2SbO_6$ film was confirmed by x-ray diffraction, $\chi(T)$, and transmission electron microscopy (Fig. S2-S4). Notably, $\chi(T)$ of bulk $Cu_3Co_2SbO_6$ shows a strong ferromagnetic transition near 60 K due to the presence of $Co_5SbO_8$ impurities.[26] In contrast, our $\chi(T)$ of the $Cu_3Co_2SbO_6$ thin film exclusively exhibits an antiferromagnetic transition near 16 K, a distinctive signature of the formation of single-phase $Cu_3Co_2SbO_6$.

**Magnetic anisotropy between bond-parallel and bond-perpendicular directions in $Cu_3Co_2SbO_6$**

For deeper understanding on the magnetic ground state of $Cu_3Co_2SbO_6$, we analyzed the magnetic anisotropy of its thin films. Despite the presence of six-fold twin domains in the thin film, the bond-parallel and bond-perpendicular directions can be uniquely determined. We measured magnetic susceptibility parallel to the bond ($\chi_{bond//}$), perpendicular to the bond ($\chi_{bond\perp}$), and perpendicular to the *ab* plane ($\chi_c$) under various *H*, as displayed in Fig. 2a-c.

Clear sharp kinks were evident in $\chi_{\text{bond}//}(T)$ and $\chi_{\text{bond}\perp}(T)$ near 16 K, while the change in $\chi_c(T)$ was relatively weaker and broader, indicating a dominant spin direction in the $ab$ plane. As $H$ increased, the kinks in $\chi_{\text{bond}//}(T)$ and $\chi_{\text{bond}\perp}(T)$ shifted to lower $T$ and became less discernible, whereas the ones in $\chi_c(T)$ remained robust even at the highest $H$. Fig. 2d-f show contour graphs of d$\chi$/d$T$ across different $H$ and $T$ for vertical and horizontal axis, respectively. These graphs visualize the paramagnetic (blue region) and antiferromagnetic (red region) phases. While in-plane $H$ significantly reduces the $T_N$, the out-of-plane $H$ hardly affects the magnetic properties. Similar observations have been reported in $\alpha$-RuCl$_3$[29,30], a well-known Kitaev QSL candidate.

In our study, we not only identified a significant anisotropy between the in-plane and out-of-plane responses but also observed a distinct anisotropy between the bond-parallel and bond-perpendicular directions. To clarify the difference between the bond-parallel and bond-perpendicular directions, we plotted $\chi_{\text{bond}//}(T)$ and $\chi_{\text{bond}\perp}(T)$ together as shown in Fig. 2g. At lower field $H$ = 10 kOe, $\chi_{\text{bond}//}(T)$ and $\chi_{\text{bond}\perp}(T)$ overlapped almost perfectly. However, as we increased the $H$, $\chi_{\text{bond}//}(T)$ and $\chi_{\text{bond}\perp}(T)$ began to deviate from each other. The upturn in low $T$ for $\chi_{\text{bond}//}(T)$ is more evident than in $\chi_{\text{bond}\perp}(T)$, suggesting the antiferromagnetic state is more fragile to $H$ in the bond-parallel direction. Fig. 2h displays the $M$-$H$ and d$M$/d$H$ curves for both directions at 2 K. The peak structures in d$M$/d$H$ typically indicate the spin-flip transition. These peaks are located near 45 kOe and 60 kOe for the bond-parallel and bond-perpendicular directions, respectively, further corroborating that the antiferromagnetic state is more vulnerable in the bond-parallel direction. It is consistent with suggested zigzag antiferromagnetic ordering in bulk where spins are aligning along the bond-perpendicular direction.[29]

The observed in-plane anisotropy of $Cu_3Co_2SbO_6$ implies its spin Hamiltonian is close to an anisotropic generalized Heisenberg-Kitaev (GHK) Hamiltonian rather than an isotropic XXZ model. The appropriate spin Hamiltonian for high-spin $3d^7$ cobalt-based materials, whether it's the XXZ or GHK model, is still a topic of debate. A key distinction between these two models is the presence of in-plane magnetic anisotropy.[15] Specifically, the GHK model has anisotropy between bond-parallel and bond-perpendicular directions in the *ab* plane,[31] while the XXZ model retains full rotation symmetry despite a slight orthorhombic distortion.[15] The evident in-plane anisotropy clearly indicates that the spin Hamiltonian of $Cu_3Co_2SbO_6$ deviates from the XXZ model. It's worth mentioning that the XXZ-$J_1$-$J_3$ model, applied to $BaCo_2(AsO_4)_2$, exhibits a negligible in-plane anisotropy in its critical magnetic fields for field-induced transitions. The different critical field between the bond-parallel and bond-perpendicular directions is more consistent with a theoretical calculation of GHK Hamiltonian.[15]

**Coupling between Exciton and Antiferromagnetic ordering of $Cu_3Co_2SbO_6$**

A unique feature of the observed antiferromagnetism in $Cu_3Co_2SbO_6$, distinguishing it from other cobalt-based honeycomb systems, is its strong interaction with excitons. Fig. 3a displays the real part of the optical conductivity $\sigma_1(\omega)$ of $Cu_3Co_2SbO_6$ film at 6 K obtained through ellipsometry. We observed a peak with anomalously high intensity near 4 eV, which has not been observed in the similar compound $Na_3Co_2SbO_6$.[32] This peak indicates an excitonic transition, similar to observations in other copper-based delafossites.[33,34] Based on the previous first-principle calculations,[35,36] we assigned the sharp 4 eV excitation to the Cu $3d \rightarrow$ Co $3d$ $e_g$ exciton. A shoulder peak around 3.6 eV can be attributed to Cu $3d \rightarrow$ Co $3d$ $t_{2g}$ transition, given that the orbital overlap between Cu $3d$ and Co $t_{2g}$ orbitals are significantly smaller than that between Cu $3d$ and Co $e_g$ orbitals (Supplementary Note1).

The $T$-dependence of the excitonic transition reveals an unconventional spin-exciton coupling, manifested not only through $T_N$ but also additionally through $T_H \sim 40$ K. Fig. 3b exhibits the raw ellipsometry parameters, $\Psi$, which represents the amplitude ratio between reflected $p$- and $s$- polarized light as a function of $T$ at exciton peak. Naively speaking, Fig. 3b reflects the $T$-dependent absorption coefficient at the exciton energy (Fig. S5). Without employing any model fitting, we identified two discernible features: a kink around $T_N \sim 16$ K and an additional anomaly near $T_H \sim 40$ K, which is more than twice of $T_N$. This unconventional kink implies a robust spin-exciton coupling that remains even above $T_N$. For a quantitative analysis, we conducted a simultaneous fitting of $\sigma_1(\omega)$ and $\varepsilon_1(\omega)$ using Lorentz-Gaussian oscillator models (Fig. 3a and Fig. S6). Fig. 3c-e display the peak position, peak intensity, and scattering rate of the excitonic transition as a function of $T$, respectively. All these fitting parameters show clear anomaly at $T_N$ through the spin-exciton coupling. In particular, the peak's blueshift begins around $T_H$, not $T_N$, consistent with the observed $T$-dependence in $\Psi$.

An anomaly in optical exciton observed at $T_H$ suggests the presence of finite spin-spin correlation functions driven by spin fluctuations, indicating strong frustration. Various researches on optical transitions in Mott and charge-transfer insulators have demonstrated that nearest-neighbor spin-spin correlations lead to spectral weight redistribution and renormalization of excitation energy.[37-40] Owing to bilinear nature of spin Hamiltonian, such spectral weight and energy renormalization is generally proportional to $|M_{sublattice}|^2$ under the mean field approximation, where $M_{sublattice}$ represents magnetization of magnetic sublattices. A comprehensive explanation on the observed peak shift below $T_N$, therefore, can be provided with increasing $M_{sublattice}$ with decreasing $T$. However, the observed peak shift between $T_N$ and $T_H$ is intriguing and beyond the mean field regime, especially given that $M_{sublattice}$ being zero above $T_N$. This shift can be attributed to the finite spin-spin correlation functions due to spin-fluctuations. Such finite spin-spin correlation functions above $T_N$ have been extensively observed in frustrated system and are considered indicative of frustration and quantum fluctuation.[41]

The observed spin-exciton coupling is further supported by peak shift and enhanced intensity under $H$ whichsuppress the antiferromagnetic ordering. Fig. 4a-b show the transmittance and absorbance spectrum of 20 nm $Cu_3Co_2SbO_6$ thin film without an applied $H$. In the absorbance spectrum, we observed a sharp peak near 4.2 eV, which is consistent with $\sigma_1(\omega)$ shown in Fig. 3a. The slight discrepancy in peak energy between $\sigma_1(\omega)$ and absorbance spectrum can be attributed to the variations in experimental method (Fig. S7). Fig. 4c-e represent the absorbance spectrum with various $H$ for each direction at 1.6 K. To offer a quantitative understanding of the changes in absorption with respect to $H$, we fitted the absorption spectrum using three Lorentzian functions (Fig. S8). Fig. 4f-h indicate the peak energy and integrated intensity between two isosbestic point (Fig. S8) for each $H$ direction. Increasing $H$ induces a distinct redshift in peak energy and increase of the intensity both in bond-parallel and bond-perpendicular directions. Given that in-plane $H$ suppresses the antiferromagnetic order, the peak energy and the intensity evolve in the opposite direction of lowering $T$. Conversely, out-of-plane $H$ leaves both the peak position and intensity unchanged. This observation corroborates the robust antiferromagnetism in that axis, as shown Fig. 2c.

The observed peak energy and intensity of the excitonic transition under $H$ also show bond-dependent anisotropy, consistent with different spin-flip transitions. The intensity of the peak, for example, begins to show an abrupt increase above 40 kOe (60 kOe) with $H$ along the bond-parallel (bond-perpendicular) direction. These fields are close to the observed spin-flip transition fields in Fig. 2h. In terms of the peak energy, we expect it will be proportional to the number of spins that flip to the ferromagnetic configuration throughout the linear Zeeman term. The flipping of a single spin can be described as the creation of spin-1 boson. Therefore, the phenomenological equation for peak energy can be expressed as follows.

$$\Delta E(H) = -\frac{A}{\exp\left(-\frac{H_C}{H}\right) - 1}$$

, where $\Delta E(H)$, $H_C$, and $A$ are the redshift of the peak, the $H$ required for spin-flip transition, and a constant, respectively. The solid lines in Fig. 4f, g represent the best fitting results with $H$ along bond-parallel and bond-perpendicular directions, respectively. Note that $H_C$ is found to be 44 kOe (66 kOe) along the bond-parallel (bond-perpendicular) direction, providing excellent coincidence with the $H$ for the spin-flip transition observed in Fig. 2h.

Figs. 3 and 4 clarify a strong coupling between antiferromagnetism and exciton, establishing $Cu_3Co_2SbO_6$ as a unique platform for the investigation and potential manipulation of quantum magnetism via light.

**Robustness of spin fluctuating region under magnetic field in $Cu_3Co_2SbO_6$**

We found that $T_H$ remains unchanged under $H$, implying the robustness of the spin-fluctuation region. Fig. 5a shows the $T$-dependent peak position obtained through absorbance under varying $H$. To discern the contrast in peak shifts on either side of $H_C$, we focus on bond-parallel directional case, which has a relatively lower $H_C$. Without an applied $H$, both the absorbance and $\sigma_1(\omega)$ reveal a similar redshift trend. This trend, however, reverses with $H$ of 70 kOe, suggesting a phase transition to a spin-polarized (SP) state and the sign change of the spin-spin correlation function. In both cases, the peak position begins to change around 30 ~ 40 K, emphasizing the robustness of $T_H$. The peak position becomes $T$-independent with applied $H$ of 40 kOe, close to the spin-flip field $H_C$, representing a balance of antiferromagnetic and ferromagnetic spin-spin correlation functions. Fig. 5b illustrates $H$-dependence of absorbance peak at a various $T$. At 10 K, given the persistence of antiferromagnetic ordering, the trend mirrors that of Fig. 4f. Notably, the presence of peak shift exists even at 20 and 30 K, where the antiferromagnetic ordering has been vanished. The peak shift become almost $H$-independent above 40 K consistent with $T_H$ of 40 K observed in $T$-dependent $\sigma_1(\omega)$ (Fig. 3c).

Although the origin of the spin fluctuating region below $T_H$ is unclear yet, one potential explanation might be a spin fractionalized region, commonly found in a Kitaev QSL system. In $\alpha$-RuCl$_3$, it is reported that the second relases of magnetic entropy and finite spin-spin correlation function at and below a characteristic temperature $T_H$, which is significantly higher than the $T_N$, respectively.[41] This spin fluctuation region below $T_H$ is understood as an unconventional Kitaev paramagnetic phase due to the occurance of spin fractionalization. In addition, recent theoretical study on the GHK model, one of the possible spin Hamiltonians for the high-spin $3d^7$ cobalt-based honeycomb materials, has resembled a similar fluctuating regime with finite spin-spin correlation functions below the conventional paramagnetic phase.[16,42] Moreover, they reported that the $T_H$ is barely changed by external $H$, which is similar to our observation in $Cu_3Co_2SbO_6$. Further studies for understanding the origin of the spin fluctuation region are highly required.

In summary, we have revealed bond-dependent antiferromagnetism, the exciton coupled to its magnetic ground state, and an unconventional spin fluctuation region between antiferromagnetic and paramagnetic phases in $Cu_3Co_2SbO_6$. The observed bond-dependent antiferromagnetism and spin fluctuation region imply that $Cu_3Co_2SbO_6$ can be a promising starting materials to realized QSL phase. Additionally, the interaction between exciton and fluctuating spin offers $Cu_3Co_2SbO_6$ with a unique platform to detect, realize, and manipulate the spin liquid using light. For example, transient spin liquid states can be achieved using a time-resolved pump-probe experiment as it perturbs its spin Hamiltonian through spin-exciton coupling.[43,44] The potential to generate and control fractional excitation via light through the exciton-spin interaction could be explored if the desired Kitaev QSL phase can be stabilized in $Cu_3Co_2SbO_6$. Lastly, we believe that our experimental approach bridges two areas prevoiusly seem as incompatible: heterostructures and quantum magnetism. While there is growing interest in applying heterostructure methods to Kitaev QSL studies,[19,45] actual implementations have been rare due to the extremely small volume available for detecting spin-spin correlation functions. However, our approach via light, unaffected by volume constraints, provides promising methodology to merge two distinct territories.

**Acknowledgement**

We thank E.-G. Moon for fruitful discussion. This work was mainly supported by the National Research Foundation (NRF) of Korea funded by the Ministry of Science and ICT(Grant No. NRF-2020R1C1C1008734), by Creative Materials Discovery Program through the National Research Foundation of Korea (NRF) funded by the Korea government (MSIT) (2017M3D1A1040834), and by the MSIT(Ministry of Science and ICT), Korea, under the ITRC(Information Technology Research Center) support program(IITP-2023-RS-2023-00259676) supervised by the IITP(Institute for Information & Communications Technology Planning & Evaluation). The Excimer Laser COMPexPro 201F (Coherent Co.) for thin film growth and M-2000 ellipsometer (J.A.Woolam Co.) for optical measurements were supported by IBS Center for Correlated Electron systems, Seoul National University. Jae Hoon Kim was supported by the Samsung Science and Technology Foundation (Grant SSTF-BA2102-04) and by the NRF of Korea (Grant No. NRF-2021R1A2C3004989 and Grant No. NRF-2017R1A5A1014862 (SRC Program: vdWMRC Center)). Jung-woo Yoo was supported by the NRF of Korea (Grant No. 2021R1A2C1008431). Yongsoo Yang was supported by the NRF of Korea (Grant No. RS-2023-00208179). The HAADF-STEM sample preparation and imaging were conducted using a FEI Helios G4 focused ion beam equipment and a double Cs corrected Titan cubed G2 60-300 (FEI) microscope at KAIST Analysis Center for Research Advancement (KARA). Excellent support by Hyung Bin Bae, Jin-Seok Choi, Tae Woo Lee and the staff of KARA is gratefully acknowledged. Yongsoo Yang also acknowledges the support from the KAIST singularity professor program and the KAIST Quantum Research Core Facility Center. (KBSI-NFEC grant funded by Korea government MSIT, PG2022004-09).

**Contributions**

B.K., U.C., and C.S. conceptualized this work. U.C., G.-H.K., U.S., M.P., J.-H.C., N.X.D., T.H.K., and C.S. synthesized and characterized the thin films. M.K., G.J., and J.M.O. synthesized a polycrystalline target for pulsed laser deposition. H.J., S.H., and Y.Y. performed transmission electron microscopy. B.K., S.N., S.S., S.P., J.-W.Y., and C.S. performed the magnetic property experiments. B.K. and U.C. performed spectroscopic ellipsometry. T.S.J. and J.H.K conducted the transmittance under the magnetic field. B.K., U.C., T.H.K., Y.Y., J.M.O., J.-W.Y., J.H.K., and C.S. analyzed the experimental data. B.K., U.C., and C.S. wrote the paper with input from all coauthors.

## Methods

### Sample preparation

High-quality $Cu_3Co_2SbO_6$ thin films were synthesized using pulsed laser deposition. The O-faced ZnO [0001] and $MgAl_2O_4$ [111] substrates were annealed for 2 hours at 1100 °C and for 10 hours at 1200 °C in ambient pressure, respectively, to improve the surface roughness and crystal quality. We made a target followed the previously reported synthesis of polycrystalline $Cu_3Co_2SbO_6$ powder using the solid-state reaction method.[26] The base pressure remained under $1\times10^{-6}$ Torr. The optimized growth conditions were as follows: substrate temperature $T = 800$ °C, oxygen partial pressure $P = 10$ mTorr, energy of the KrF Excimer laser ($\lambda = 248$ nm) $E = 1.3$ J/cm$^2$, laser repetition rate = 10 Hz, and the distance between the target and substrate was set at 50 mm. Cooling was performed under the same as grown pressure after the deposition was completed. No sample degradation was observed, even when the target and the synthesized thin film were stored at room temperature and under ambient pressure.

**Characterization of lattice structure and film thickness**

By using a D8 Discovery high-resolution X-ray diffraction (Bruker), high resolution X-ray diffraction data of $Cu_3Co_2SbO_6$ thin film were collected at room temperature using a wavelength of 1.5406 Å. Using the 0D mode of Lynxeye detector, $\theta$-$2\theta$ scan was performed at 0.005° intervals from 10° to 80° in $2\theta$ and at a scan speed of 0.016° s–1. The rocking curves were performed at 0.005° intervals from ± 1.5° of 004 peak of $Cu_3Co_2SbO_6$ with 0.5 step/sec. An azimuthal $\varphi$-scans of $Cu_3Co_2SbO_6$ thin film were conducted at angle $\chi$ = 54.7356° with respect to ZnO (0001) surface. Collected angle range is from -210° to 150° with 0.02° increments and 0.5 step/sec scan speed for ZnO substrate and $Cu_3Co_2SbO_6$ thin film. Note that counterclockwise rotation is positive.

**Magnetic susceptibility**

The temperature dependence of the zero-field-cooled and field-cooled d.c. magnetization of 14.2 mg (13.6 mg) $Cu_3Co_2SbO_6$/ZnO ($Cu_3Co_2SbO_6$/$MgAl_2O_4$) thin film, with dimension of 3.3 mm (bond-parallel) x 2.5 mm (bond-perpendicular) x 0.33 mm (0.5 mm for $MgAl_2O_4$) (out-of-plane), was measured in each direction using a superconducting quantum interference device (Quantum Design). The measurements were conducted by attaching the samples to the quartz sample mounting post with GE varnish. Similarly, the magnetic properties of 15.2 mg (13.6 mg) bare ZnO ($MgAl_2O_4$), of the same dimension, were obtained. The magnetization of $Cu_3Co_2SbO_6$ was calculated by subtracting the mass-normalized substrate magnetic susceptibility. To compare relatively small difference in $\chi_{bond//}(T)$ and $\chi_{bond\perp}(T)$, temperature- and field-independent background contribution from environment is further subtracted based on high-temperature Curie-Weiss fitting. This process is valid with the known fact that bulk $Cu_3Co_2SbO_6$ has a negligible constant background contribution $\chi_0$ in magnetic susceptibility.[26]

**Scanning transmission electron microscopy measurements**

For transmission electron microscopy (TEM) analysis, a cross-section specimen of approximately 50 nm was fabricated using a focused ion beam (FIB) machine (Helios G4, FEI) from a 20 nm $Cu_3Co_2SbO_6$ film grown on a ZnO substrate. The interface structure was then measured using a Titan Double Cs corrected TEM (Titan cubed G2 60–300, FEI) with high-angle annular dark field scanning TEM (HAADF-STEM) mode. The microscope was operated at 300 kV accelerating voltage with the beam convergence semi-angle of 25.2 mrad. The inner and outer angles of the HADDF detector were set as 38 and 200 mrad, respectively. Several 1024 × 1024 images of the film-substrate interface were acquired with 4 μs dwell time per pixel. The pixel size was 6.48 pm for Fig. S2e, 12.96 pm for Fig. S3a-b, and 25.91 pm for Fig. S3c. The total electron dose was about $4.64 \times 10^5$ electrons $Å^{-2}$ for Fig. S2e, $1.18 \times 10^5$ electrons $Å^{-2}$ for Fig. S3a, $1.13 \times 10^5$ electrons $Å^{-2}$ for Fig. S3b, and $2.94 \times 10^4$ electrons $Å^{-2}$ for Fig. S3c.

**Ellipsometry and optical conductivity**

The optical conductivity of 20 nm $Cu_3Co_2SbO_6$ was obtained using an M-2000 ellipsometer (J. A. Woolam Co.). The ellipsometry parameters, $\Psi$ and $\Delta$, of $Cu_3Co_2SbO_6/MgAl_2O_4$ were measured over the energy range of 0.74 to 6.46 eV (5900 to 52000 $cm^{-1}$) at 60° incident angles and temperatures (6 to 300 K). Each measurement had a duration of 200 seconds. Here, $\Psi$ represents the amplitude ratio of the reflected *p*- and *s*- waves, while Δ represents the phase shift between the two waves. We also obtained Ψ and Δ of the substrate under the same incident angles and temperatures. We determined the optical constants of $Cu_3Co_2SbO_6$ layers by constructing thin film models including intermixing layer and surface roughness. For low-temperature measurement, we calibrated the window effect to determine the Δ offset by using a 25 nm $SiO_2$/Si wafer. To prevent ice formation on the sample surface, we baked out a chamber to obtain the base pressure below $5 \times 10^{-9}$ Torr. All samples were attached with silver epoxy to oxygen-free copper cones to prevent reflections from the backside of the sample.

**Transmittance under applied magnetic field**

The transmittance spectra of $Cu_3Co_2SbO_6$ thin film on $MgAl_2O_4$ substrate were measured using a Deuterium light source (SLS204, Thorlabs) and a charge-coupled device (CCD) spectrometer (CCS200, Thorlabs). To reduce background noise, a dark measurement was subtracted from all raw spectra. The transmission experiments under the external magnetic fields were conducted using a magneto-optic chamber (SpectromagPT, Oxford Instruments), which has 4 optical ports to facilitate optical experiments in both vertical and horizontal directions relative to the external magnetic fields.

[*] These authors contributed equally.
[§] jwyoo@unist.ac.kr
[‡] super@yonsei.ac.kr
[†] chsohn@unist.ac.kr

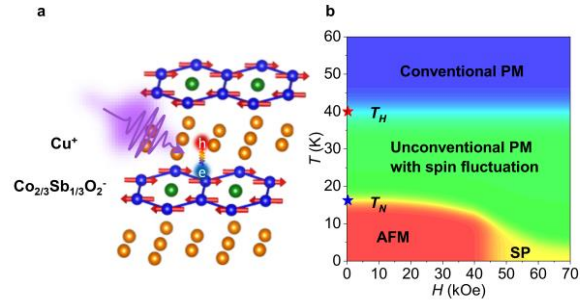

**Fig. 1: Schematics of spin-exciton coupling and the phase diagram of $Cu_3Co_2SbO_6$. a,** Layered crystal structures of $Cu_3Co_2SbO_6$ with alternative $Cu^+$ and $(Co_{2/3}Sb_{1/3})O_2^-$ layers. $(Co_{2/3}Sb_{1/3})O_2^-$ layers has two-dimensional honeycomb structures with zigzag antiferromagnetic ordering perpendicular to the Co-Co bond direction. With incident light, there is formation of excitons between $Cu^+$ and $(Co_{2/3}Sb_{1/3})O_2^-$ layers, which interact with antiferromangetism. **b,** Suggested magnetic phase diagram of $Cu_3Co_2SbO_6$ with $H$ along the bond-parallel direction (PM: paramagnetic region, AFM: antiferromagnetic region, SP: spin-polarized region). The $T_N$ and $T_H$ is obtained from the $\chi(T)$ measurement and optical spectroscopy, respectively. Unconventional spin fluctuating regime was observed between low-$T$ AFM and high-$T$ PM phases.

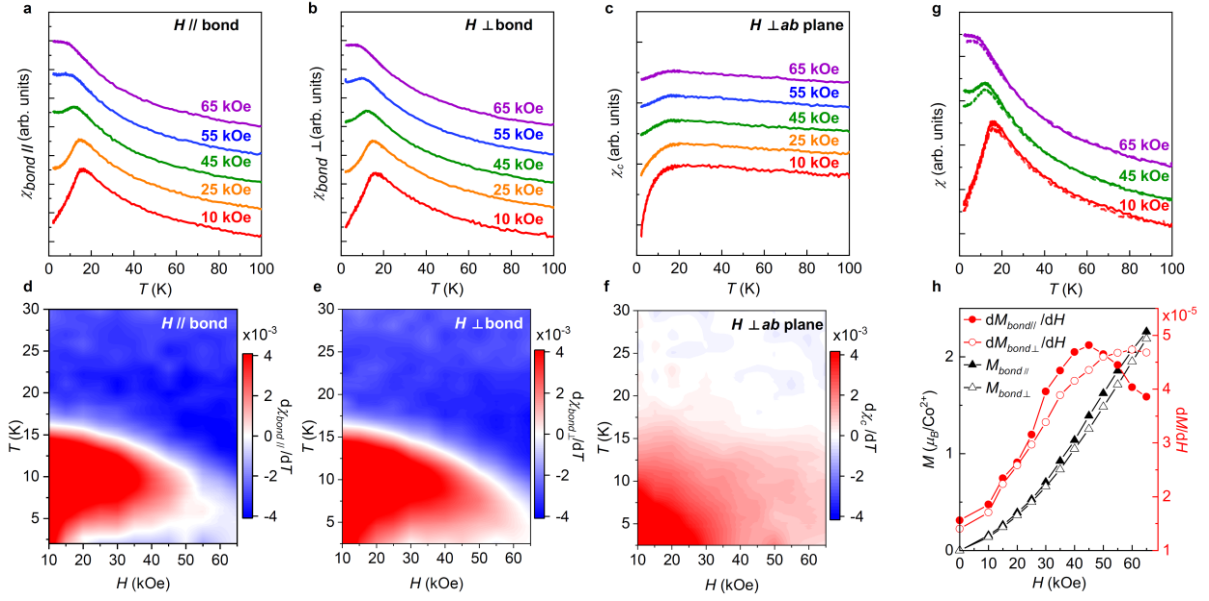

**Fig. 2: Bond-dependent antiferromagnetism in $Cu_3Co_2SbO_6$.** $\chi(T)$ of $Cu_3Co_2SbO_6$ under different orientations of applied $H$ **a**, along the bond-parallel direction, **b**, along the bond-perpendicular direction, and **c**, perpendicular to honeycomb planes. The $T$ and $H$ dependent contour plot of derivative susceptibility, $d\chi/dT$, **d**, along the bond-parallel direction, **e**, along the bond-perpendicular direction, and **f**, perpendicular to honeycomb planes. The antiferromagnetic and paramagnetic regions are indicated by red and blue areas, respectively. **g,** An overlapped graph of $\chi_{bond//}(T)$ and $\chi_{bond\perp}(T)$ presenting in-palne magnetic anisotropy under various $H$ **h,** $M$-$H$ and $dM/dH$ curves with $H$ along the bond-parallel and bond-perpendicular directions. Open and closed triangles (circles) indicate the bond-parallel and bond-perpendicular directions $M$-$H$ ($dM/dH$) curves, respectively.

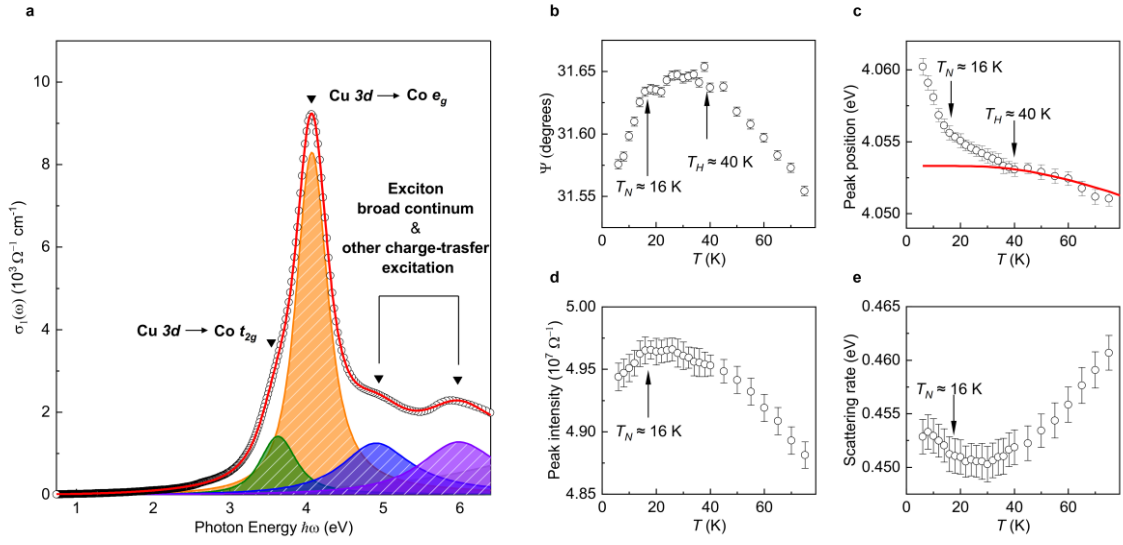

**Fig. 3: Optical exciton and spin-exciton coupling in $Cu_3Co_2SbO_6$. a,** $\sigma_1(\omega)$ at 6 K (open circles) and Lorentz-Gaussian oscillator fitting (solid lines) for $Cu_3Co_2SbO_6$. The strong excitonic excitation, attributed to the Cu 3d → Co 3d $e_g$ transition, is located near 4.1 eV. The *T*-dependent **b,** the ellipsometry data Ψ **c,** exciton peak position, **d,** peak intensity, and **e,** scattering rate. The red line in **c** is Bose-Einstein statistics fitting functions implemented at high *T* data, generically found in spectra due to phonon contribution. The black arrows indicate the $T_N$ and $T_H$. The distinct kink below $T_N$ in all fitting parameters implies the existence of coupling between the exciton and antiferromagnetic ordering. The clear additional anomaly in Ψ and peak position at $T_H$ indicates the presence of finite spin-spin correlation functions above $T_N$.

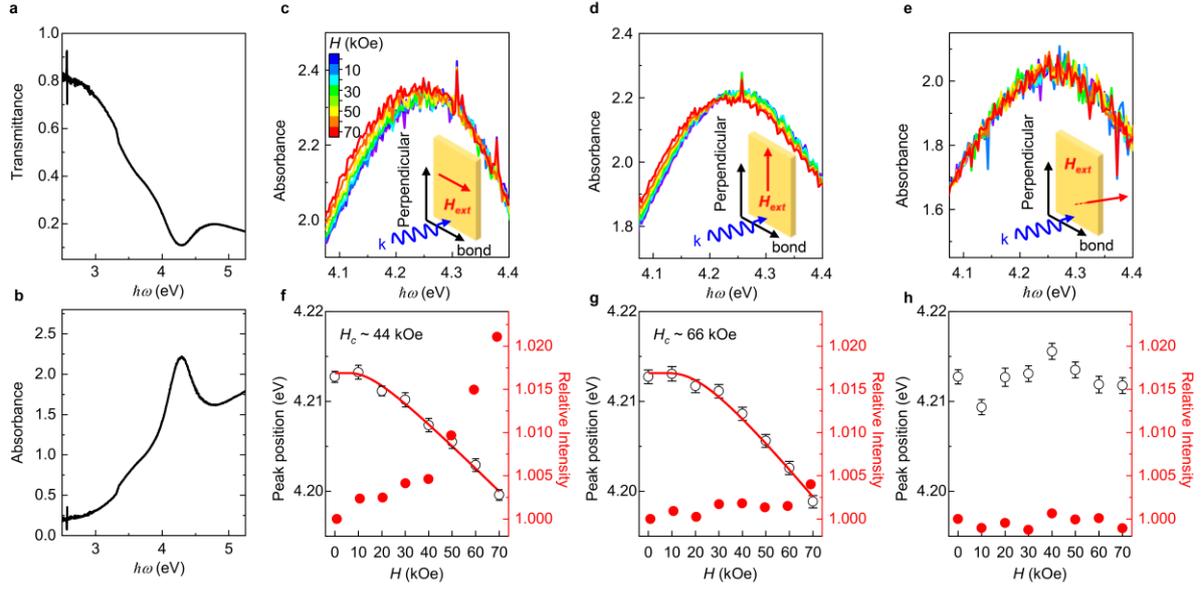

**Fig. 4: Modulation of exciton peak energy and intensity driven by external magnetic field $H$. a,** Transmittance and **b,** absorbance of $Cu_3Co_2SbO_6$ without applied $H$. The absorbance of $Cu_3Co_2SbO_6$ under various applied $H$ **c,** along the bond-parallel direction, **d,** along the bond-perpendicular direction, and **e,** perpendicular to honeycomb planes. Red and blue arrows represent the direction of applied $H$ and incident light, respectively. The redshift of absorbance in in-plane $H$ is attributed to the suppression of antiferromagnetic ordering. Conversely, the invariable peak position observed in **e** is consistent with the robustness of antiferromagnetism against out-of-plane $H$. The exciton peak position and intensity as a function of $H$ **f,** along the bond-parallel direction, **g,** along the bond-perpendicular direction, and **h,** perpendicular to honeycomb planes. The black open circles, red closed circles, and red lines indicate the peak position, intensity, and fitting function for the peak position, respectively. The fitting function provides the critical $H_C$, consistent with observed spin-flip transition.

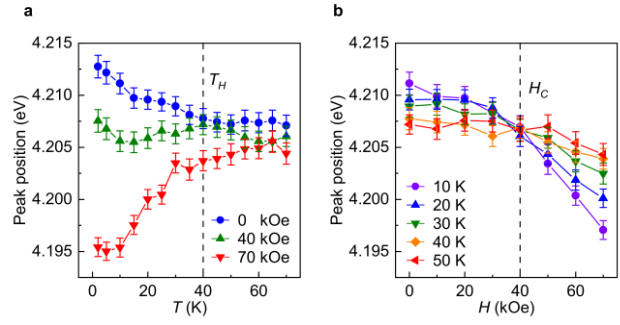

**Fig. 5: The robustness of $T_H$ and spin-fluctuation region under $H$. a,** $T$-dependent exciton peak position under different $H$. The different trends of 0 kOe, 40 kOe, and 70 kOe data reflect the antiferromagnetic, intermediate, and ferromagnetic phase, respectively. While the sign of spin-spin correlation function has changed with applied $H$, the peak shift emerges around similar $T_H$ in both $H = 0$ and 70 kOe cases. **b,** $H$-dependent exciton peak position under various $T$. The peak shift at 10 K, where antiferromagnetic ordering remains, shows a similar behavior at 2 K. Despite the disappearance of antiferromagnetic ordering at 20 K and 30 K, the persistent peak shift suggests finite spin-spin correlation functions. The disappearance of the peak shift at 40 K, again consistent with $T$-dependent peak shift in Fig. 3c.

Optical detection of bond-dependent and frustrated spin in the two-dimensional cobalt-based honeycomb antiferromagnet $Cu_3Co_2SbO_6$

# Contents

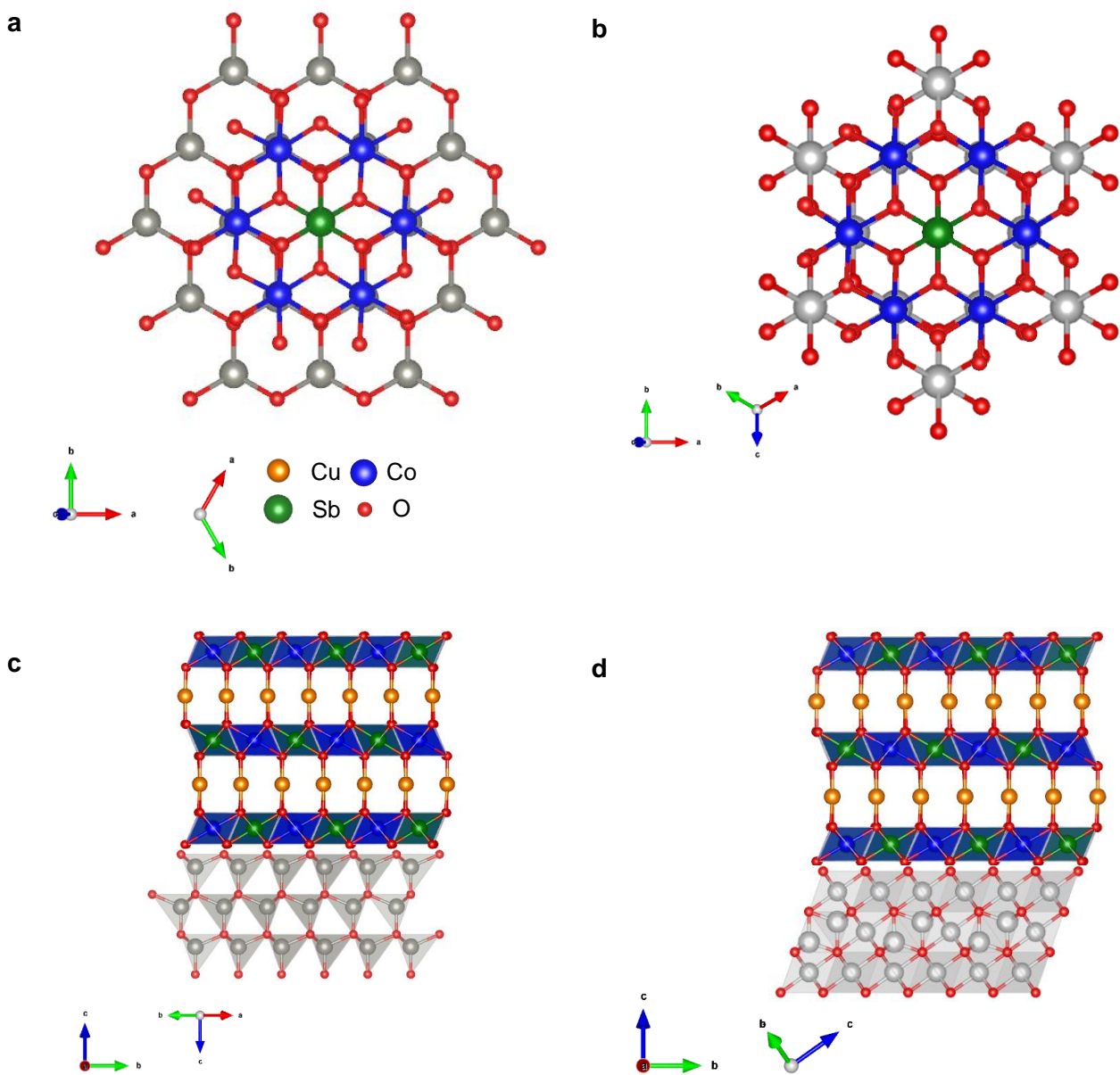

**Figure S1:** Top view of interface in **a,** $Cu_3Co_2SbO_6$[001]/ZnO[0001] and **b,** $Cu_3Co_2SbO_6$[001]/$MgAl_2O_4$ [111]. The cations of substrates are indicated as grey circle. In both case, $Cu_3Co_2SbO_6$ share the same oxygen position with substrates, promoting the epitaxial growth and high crystal quality. A side view of interface in **c,** $Cu_3Co_2SbO_6$[001]/ZnO[0001] and **d,** $Cu_3Co_2SbO_6$[001]/$MgAl_2O_4$ [111]. The shared oxygen locations and the orientation of the cobalt hexagon at the interface were determined through the azimuthal $\varphi$-scan and transmission electron microscopy.

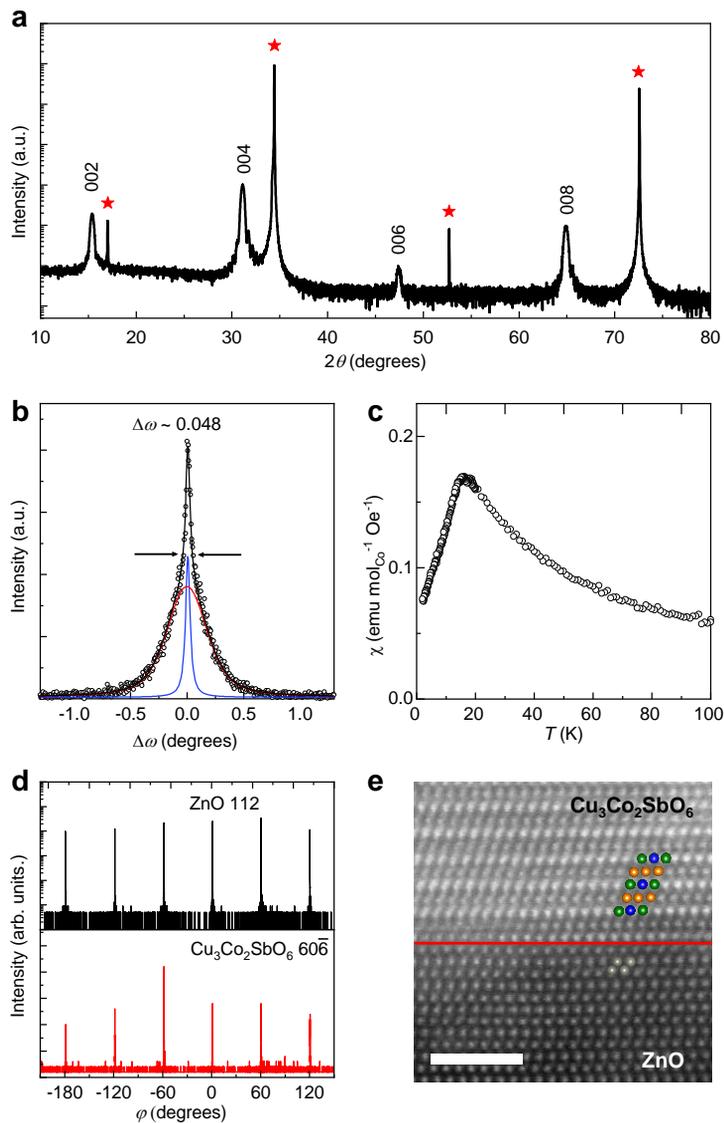

**Figure S2: a,** X-ray diffraction (XRD) $\theta$-$2\theta$ data of $Cu_3Co_2SbO_6$ thin film grown on ZnO (0001) substrate. Red stars indicate the 000$l$ peaks of ZnO. The peak positions are consistent with the reported one **b,** A rocking curve of (004) peak of $Cu_3Co_2SbO_6$. Black open circles and solid lines represent the experimental data and Lorentzian fitting functions, respectively. The sharp full width at half maximum of the blue solid line suggests high crystallinity. **c,** $\chi(T)$ of $Cu_3Co_2SbO_6$ at a $H = 10$ kOe, showing antiferromagnetic ordering near 16 K. **d,** Azimuthal $\varphi$-scans of the $(60\bar{6})$ peak of $Cu_3Co_2SbO_6$ and (112) peak of ZnO, establishing epitaxial relationship. While six-fold symmetry of (112) peak of ZnO is attributed to six-fold lattice symmetry of ZnO, the six-fold $(60\bar{6})$ peak of $Cu_3Co_2SbO_6$ originate from the presence of equivalent twin domains. **e.** A high-angle annular dark field scanning transmission electron microscopy (HAADF-STEM) image of $Cu_3Co_2SbO_6$/ZnO film. The red line and gray circles mark the interface and Zn atoms, respectively. The scale bar is 2 nm.

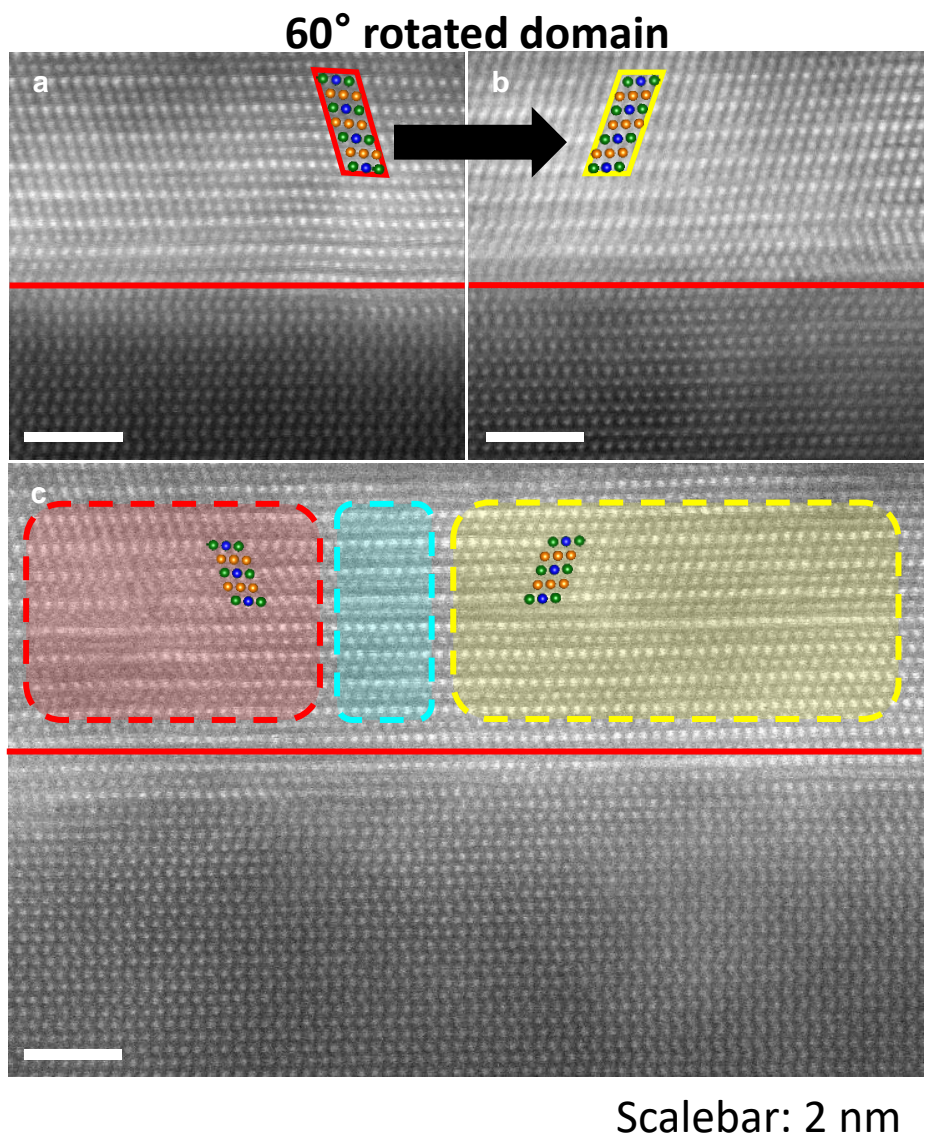

Scalebar: 2 nm

**Figure S3: a,b** HAADF images of a $Cu_3Co_2SbO_6$ thin film grown on a ZnO substrate for two different regions. The red and yellow boxes highlight the different local structures of domains related with 60° in-plane rotation. Schematic side views of the atomic structures for each domain are overlaid for reference. **c,** The region where three different domains co-exist. The red, yellow and cyan colored boxes represent a domain, a domain related with 60° in-plane rotation relative to the domain of red box, and the boundary between the two different domains, respectively. Each domain can be attributed to equivalent twin domain of $Cu_3Co_2SbO_6$. The red solid line indicates the interface between the $Cu_3Co_2SbO_6$ thin film and ZnO substrate.

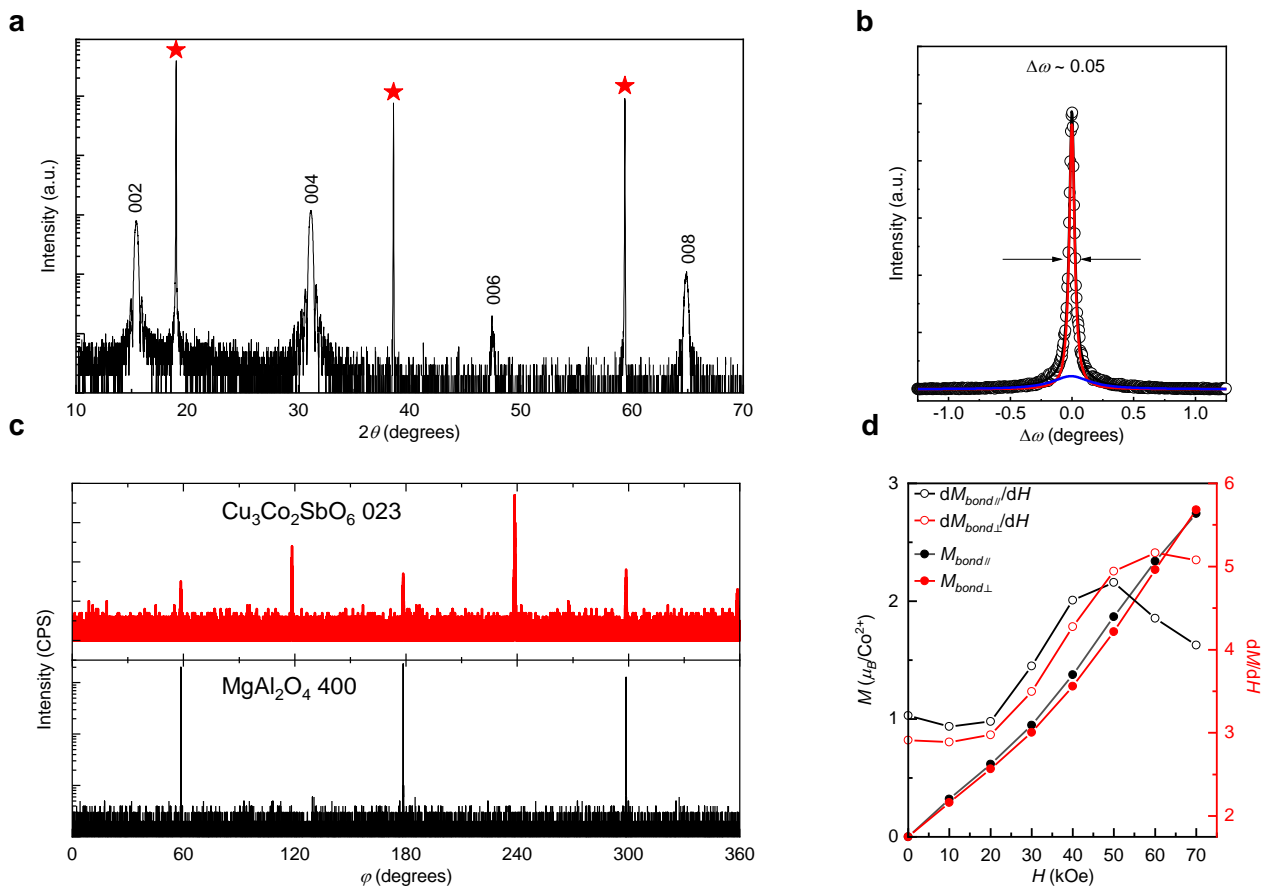

**Figure S4: a,b** XRD $\theta$-$2\theta$ of the (00$l$) peak of $Cu_3Co_2SbO_6$ thin film on $MgAl_2O_4$ [111] substrates. The red stars denote the ($lll$) peaks of $MgAl_2O_4$. **b,** The rocking curve of (004) peak of $Cu_3Co_2SbO_6$. The open circles and solid lines are corresponding to experimental data and the Lorentzian fitting functions, respectively. **c**, The azimuthal $\varphi$-scan of $Cu_3Co_2SbO_6$ (023) peak (top) and $MgAl_2O_4$ (400) peak (bottom). The observed multidomain structure is consistent with that of $Cu_3Co_2SbO_6$ on ZnO. **d,** $M$-$H$ and d$M$/d$H$ curves of $Cu_3Co_2SbO_6$ on $MgAl_2O_4$ substrate along bond-parallel direction and bond-perpendicular direction. Black and red closed (open) circles indicate the bond-parallel and bond-perpendicular direction in $M$-$H$ (d$M$/d$H$) curves, respectively. All data consistently indicate the structural and magnetic properties of $Cu_3Co_2SbO_6$ remain unaffected by the choice of substrate.

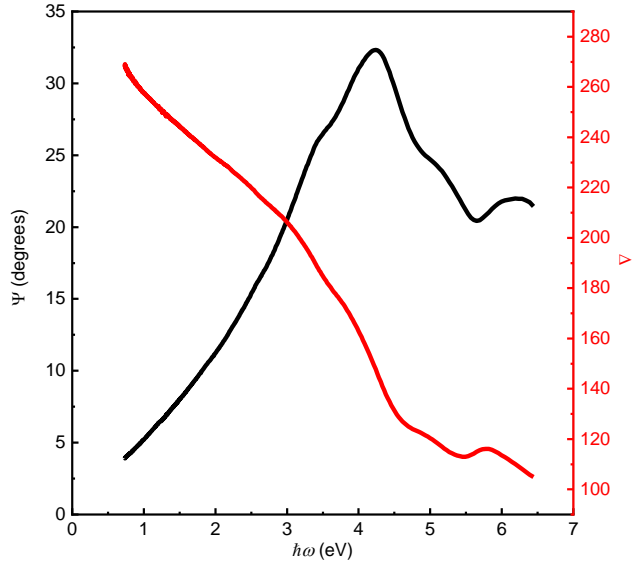

**Figure S5:** The ellipsometry parameters, Ψ and Δ, for $Cu_3Co_2SbO_6$ on $MgAl_2O_4$ at 6 K with an incident angle of 60 degrees. Ψ denotes the amplitude ratio of the reflected *p*-wave and *s*-wave, whereas Δ represents the phase shift between the two waves. Without appropriate modeling, ellipsometry parameters generally lack a clear physical interpretation, as it can be influenced by the various factors such as the incident angle of the light or the thickness and surface roughness of film. To extract quantitative physical information, we need to transform these data into optical constants like the dielectric function, optical conductivity, and refractive index with appropriate modelling process. Nonetheless, it is important to note that ellipsometry parameters and optical constants are not independent quantities; any changes in the optical constants inevitably influence the ellipsometry parameters. Given that the trend of Ψ closely resembles the imaginary part of dielectric function $\varepsilon(\omega)$, we can utilize the *T*-evolution of Ψ to extract the $T_N$.

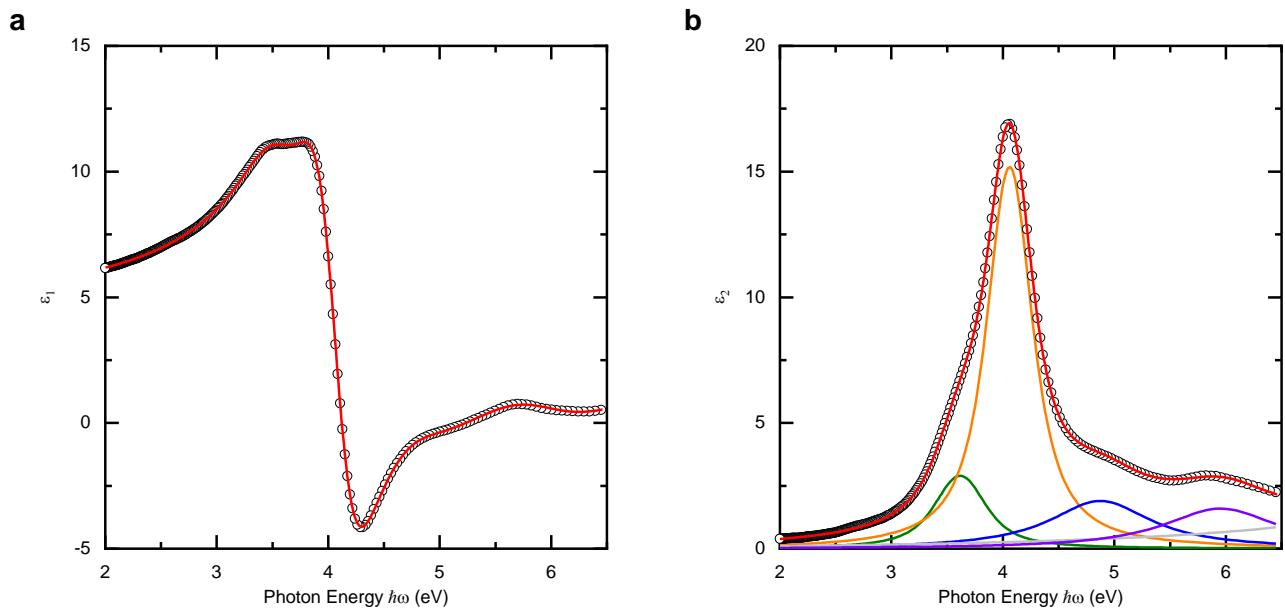

**Figure S6: a,** The real and **b**, imaginary part of dielectric function $\varepsilon(\omega)$ of $Cu_3Co_2SbO_6$ at 6 K. The open circles represent experimental data, while solid lines shows Lorentzian-Gaussian fitting function lines. Note that due to the Kramers-Kronig relation, the real and imaginary parts of the optical constants are not independent parameters, necessitating that we simultaneously fit $\varepsilon_1(\omega)$ and $\sigma_1(\omega) = \omega\varepsilon_2(\omega)/4\pi$.

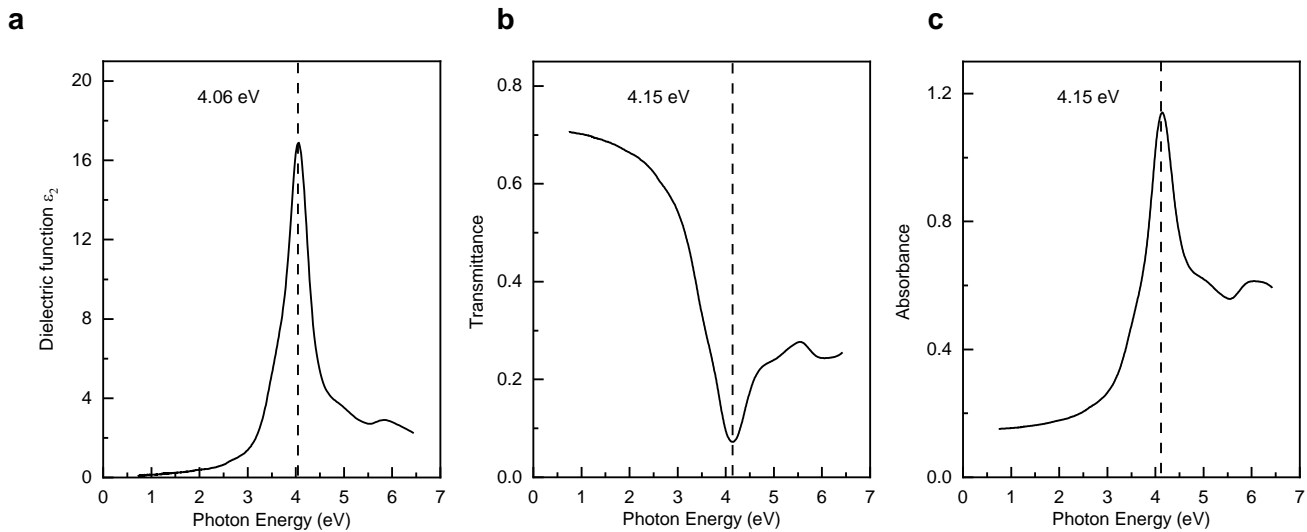

**Figure S7: a,** The imaginary part of dielectric function $\varepsilon_2$ of $Cu_3Co_2SbO_6$ at 6 K. **b,** The simulated transmittance and **c**, the simulated absorbance of $Cu_3Co_2SbO_6$. The dashed lines indicates the peak position. The different peak position between $\varepsilon_2$ and absorbance simply originates from different ways of showing optical spectra. This discrepancy originates from the assumption of the zero reflectance during transforming transmittance(T) into absorbance $(A = Log(1/T))$

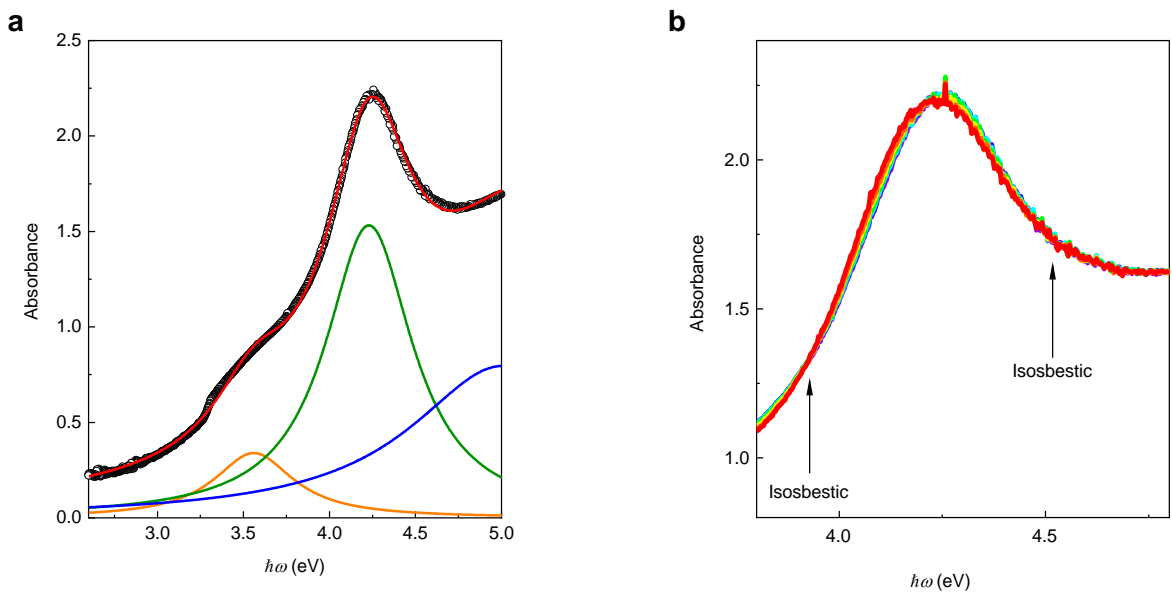

**Figure S8: a,** Lorentz fitting of absorbance spectrum. Open circles and solid lines are experimental data and Lorentz fitting functions, respectively. **b,** Isosbestic points of absorbance spectrum. To determine precise isosbestic points, we subtract of 70 kOe data to 0 kOe data. To obtain the intensity, we integral the absorbance spectrum between two isosbestic points.

# Supplementary Note 1 : Hopping integral calculation for Cu 3d to Co 3d optical transition

In order to compare the possible hopping path between $e_g^\pi/e_g^\sigma$ orbital and out-of-plane O-Cu-O dumbbell structure, the hoping amplitude for Co-O-Cu is calculated. The configuration of the $CoO_6$ system is shown in Fig. S9.

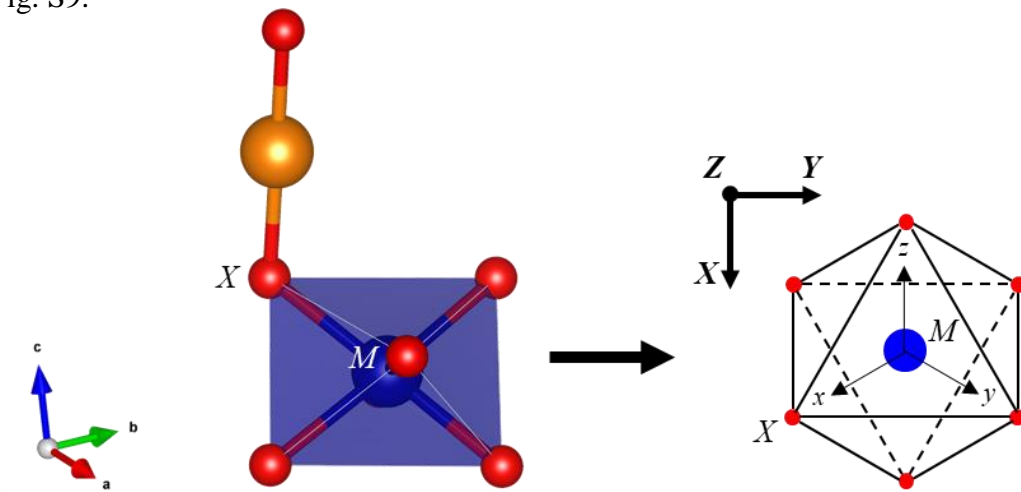

**Figure S9: Left**, the schematic of $CoO_6$ octahedra and O-Cu-O dumbbell structure of $Cu_3Co_2SbO_6$. The colors of atom is same with Fig. S1. The $Co^{2+}$ ion is in M site and O-Cu-O dumbbell structure is on X site, along the out-of-plane direction mostly. **Right,** the schematic of global (Upper bold arrows) and local coordination (Lower arrows) of $CoO_6$ octahedra.

In local octahedra coordinate of $CoO_6$ (see Figure S9), $e_g^\pi$, and $e_g^\sigma$ orbitals of $CoO_6$ can be expressed below[1]

$$e_{g,1}^\pi = -\frac{1}{\sqrt{3}}(d_{xy} + \omega d_{yz} + \omega^2 d_{zx})$$

$$e_{g,2}^\pi = \frac{1}{\sqrt{3}}(d_{xy} + \omega^{-1} d_{yz} + \omega^{-2} d_{zx})$$

$$e_{g,1}^\sigma = -\frac{1}{\sqrt{2}}(d_{3z^2-r^2} + id_{x^2-y^2})$$

$$e_{g,2}^\sigma = \frac{1}{\sqrt{2}}(d_{3z^2-r^2} - id_{x^2-y^2})$$

We considered the case of zero trigonal field for simplicity. In that case, p-d hopping matrix between d orbital at M site and p orbital in X site can be described as,

$$T_{MX}^{dp} = \begin{pmatrix} \frac{\sqrt{3}}{2}V_{pd\sigma} & 0 & 0 \\ -\frac{1}{2}V_{pd\sigma} & 0 & 0 \\ 0 & 0 & 0 \\ 0 & 0 & V_{pd\pi} \\ 0 & V_{pd\pi} & 0 \end{pmatrix}$$

in basis of $(d_{x^2-y^2}, d_{3z^2-r^2}, d_{yz}, d_{zx}, d_{xy})^\dagger$ and $(p_x, p_y, p_z)$.[2]

Considering that the topmost valence band is mostly composed of Cu $d_{3z^2-r^2}$ orbital due to crystal field from the dumbbell structure, the lowest optical excitation is from Cu 3d orbital to Co 3d orbital through hybridization between Cu 3d and O 2p orbital. From the dumbbell O-Cu-O structure, the out-of-plane $d_{3z^2-r^2}$ orbital of Cu electron only hybridizes with oxygen $p$ orbitals oriented in the (111) direction in the local octahedral coordinate system.

$$\left| p_{(1,1,1)} \right\rangle = \frac{1}{\sqrt{3}}(p_x + p_y + p_z)$$

Therefore, intensity of Cu 3d to Co 3d orbitals solely depends on the hopping matrix between Co d orbitals to the O $p_{(1,1,1)}$ orbital. Following the hopping matrix analysis, we estimated the hopping amplitude for $e_g^\pi$ and $e_g^\sigma$ orbitals.

$$\left| \langle e_g^\pi | T_{MX}^{dp} | p_{(1,1,1)} \rangle \right|^2 = \frac{V_{pd\pi}^2}{9}$$

$$\left| \langle e_g^\sigma | T_{MX}^{dp} | p_{(1,1,1)} \rangle \right|^2 = \frac{V_{pd\sigma}^2}{6}$$

Using the empirical relation $V_{pd\sigma} = -\frac{4}{\sqrt{3}} V_{pd\pi}$, it is confirmed that within the same out-of-plane geometry, the probability of optical transition from Co orbitals to Cu orbital is about 8 times higher for the $e_g^\sigma$ orbital compared to the $e_g^\pi$ orbital.